\documentstyle[aps,twocolumn,prb]{revtex}

\begin{document}
\draft \wideabs{\title{Antiferromagnetic Alignment and Relaxation
Rate of Gd Spins in
 the High Temperature Superconductor GdBa$_2$Cu$_3$O$_{7-\delta}$}
\author{R. J. Ormeno and  C. E. Gough}
 \address{School of Physics and Astronomy, The University of Birmingham,
 Edgbaston, Birmingham,
 B15 2TT, United Kingdom.}
\author{Guang Yang}
\address{Department of Metallurgy and Materials, The University of
Birmingham, Birmingham, B15 2TT, United Kingdom.}
 \maketitle
\begin{abstract}
The complex surface impedance of a number of
GdBa$_2$Cu$_3$O$_{7-\delta}$ single crystals has been measured at
10, 15 and 21 GHz using a cavity perturbation technique. At low
temperatures a  marked increase in the  effective penetration
depth and surface resistance is observed associated with the
paramagnetic and antiferromagnetic alignment of the Gd spins. The
effective  penetration depth has a sharp change in slope at the
N\'eel temperature, $T_N$, and the surface resistance     peaks
at a frequency dependent temperature below 3K. The observed
temperature and frequency dependence can be described by a model
which assumes a negligibly small interaction between the Gd spins
and the electrons in the superconducting state, with a frequency
dependent magnetic susceptibility and a Gd spin relaxation time
$\tau_s $ being a strong function of temperature. Above $T_N$,
$\tau_s$ has a component varying as $1 / (T - T_N)$, while below
$T_N$ it increases $\sim T^{-5}$.

\end{abstract}
}

 One of the surprising
early discoveries about high temperature superconductors was
their apparent insensitivity to out-of-plane magnetic ions, with
the superconducting properties of YBCO remaining almost unchanged
when yttrium was replaced by magnetic rare earth ions, with the
exception of Ce, Pr and Tb \cite{Hor}. Furthermore,  the low
temperature antiferromagnetic properties were observed to be
independent of doping. The thermodynamic superconducting
 and antiferromagnetic properties of  GdBa$_2$Cu$_3$O$_{7-\delta}$ (GBCO)
  therefore appear to be completely uncoupled \cite{dunlap,Mook}. However,
there remains the possibility that the mutual interaction of the
rare earth (RE) spins and the superconducting electrons could
lead to changes in their dynamic properties.
 Interest in the antiferromagnetic alignment of the rare earth spins
 has also
re-emerged as a likely explanation  of the anomalous increase in
the microwave surface resistance of GBCO thin films observed at
low temperatures \cite{Waldram,Anlage,Wingfield2}.

We have therefore undertaken a systematic microwave investigation
of the antiferromagnetic spin alignment in  a number of GBCO
single crystals. The surface impedance  has been measured to well
below the Gd N\'eel temperature $T_N \sim 2.2$K, using a hollow
dielectric resonator technique \cite{Wingfield}. Measurements at
10, 15 and 21 GHz confirm the influence on microwave properties
of the alignment of the Gd spins and enable us to determine the
Gd spin relaxation time $\tau_s$ both above and below the N\'eel
temperature.

In the superconducting state the microwave surface impedance is
given by
\begin{equation}
Z_s =\sqrt{\frac{{\rm i}\mu_r\mu_0\omega}{(\sigma_1-{\rm
i}\sigma_2)}}, \label{Zs}
\end{equation}
 where $\sigma_1$ is the normal state
quasi-particle conductance $n_{qp}e^2\tau_{qp}/m$ ($\tau_{qp}$ is
the quasi-particle scattering time).   In the ideal
superconducting state $\sigma_2 (T)= n_s (T) e^2/m\omega=
(\lambda (T)^2 \mu_0\omega)^{-1}$, where $\lambda (T)$ is the
penetration depth unperturbed by any coexistent magnetic
properties and  $n_s$ is the    superfluid density. However, in
the presence of magnetic spins $\sigma_2= (\lambda^2
\mu_r\mu_0\omega)^{-1}$ and we can assume a relaxation model for
a frequency dependent permeability, $\mu_r (T, \omega)= 1 + \chi
(T) /(1 +{\rm i}\omega \tau_s (T))$, with a spin lattice
relaxation time $\tau_s$. In this situation the field penetration
is modified but not the intrinsic penetration depth which is
related to the superfluid fraction. For a antiferromagnetic
system above the N\'eel temperature $T_N$ we expect $\chi (T)=
C/(T + T_N)$, with $C=n\mu_B^2p^2/12\pi k_B$, where $n$ is the
number of spins per unit volume.
 DC magnetic measurements give  an  effective moment \cite{Thompson}
$p=7.82$ , in good agreement with the free ion value of 7.92.
Rewriting $Z_s$ as $[{\rm i}\mu_0\omega/\sigma_{eff}]^{1/2}$, we
can define an effective conductivity  as
\begin{equation}
 \sigma_{eff}= \frac{(\sigma_1-{\rm
i}\sigma_2)}{1 + \chi/(1+{\rm i}\omega\tau_s)}.
\end{equation}
The real and imaginary parts are give by
\begin{equation}
\sigma_{1eff} = \frac{(1+\omega^2\tau_s^2)(\Gamma\sigma_1 +
\sigma_2\chi\omega\tau_s)}{\Gamma^2+\chi^2\omega^2\tau_s^2},
\label{fullsigma1}
\end{equation}
and
\begin{equation}
\sigma_{2eff} =
\frac{(1+\omega^2\tau_s^2)(\sigma_1\chi\omega\tau_s -
\Gamma\sigma_2)}{\Gamma^2+\chi^2\omega^2\tau_s^2},
\label{fullsigma2}
\end{equation}
where $\Gamma = 1+\omega^2\tau^2_s +\chi$.
 At low temperature, we will show that $\omega\tau_s>1$, so
that
\begin{equation}
\sigma_{1eff}\sim \sigma_1 + \sigma_2\chi/\omega\tau_s
\end{equation}
 and
 \begin{equation}
\sigma_{2eff}\sim \sigma_2(1 +
\chi/\omega^2\tau^2_s+2/\omega^2\tau_s^2 ) -
\sigma_1\chi/\omega\tau_s, \end{equation}
 where we retained
terms to second order in  $1/\omega\tau_s$ because
$\sigma_2\gg\sigma_1$.

In fitting our data we have also assumed  no change in the
quasi-particle conductance from the Gd spin fluctuations. To test
the validity of this model, we have measured the surface
impedance of several GBCO single crystals both above and below
$T_N$ and at several microwave frequencies.

\begin{figure}[tbh]
\begin{center}
\input epsf
\epsfxsize=85mm  \epsfbox{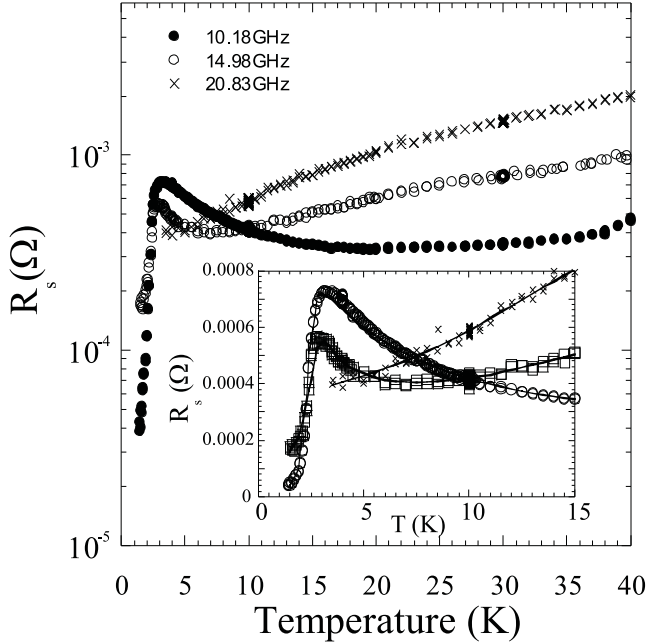} \caption{The surface
resistance $R_s (T)$ for currents flowing in the $a-b$ plane of a
high quality GdBa$_2$Cu$_3$O$_{7-\delta}$ single crystal. The
inset shows an expanded view of the low temperature region. The
solid lines represent calculated values of $R_s (T)$ using
 equation \ref{fullsigma1}, with  values of $\chi (T)$ and
$\tau_s$ derived from our model.} \label{Rs}
\end{center}
\end{figure}
The GBCO crystals were grown in BaZrO$_3$ crucibles to minimise
contamination \cite{Liang,Erb}.The largest crystal used in these
measurements was $1.0 \times 1.1 \times 0.06 {\rm mm^3}$ with a
$T_c$  of 93K. During the course of the experiments, with the
sample held overnight at room temperature under vacuum, the
microwave transition at $T_c$  was observed to broaden, with a
second transition  emerging  at $\sim 63$K. This second
transition is  similar to earlier measurements by Srikanth {\it
et al.} on a similarly grown YBCO crystal, which was  interpreted
as a second energy gap \cite{twogap}. We believe the anomaly is
more likely to be associated with oxygen diffusion  out of the
sample, resulting in surface  regions of oxygen deficient, 60K
phase. This may be a generic problem for BaZrO$_3$-grown HTS
crystals held in vacuum at room temperature for any length in
time. However, no associated changes were observed in the low
temperature microwave properties in our measurements.

The surface impedance was measured using a cylindrical dielectric
resonator with a 2mm hole passing along its axis. The resonator
was placed centrally within an OFHC copper cavity. Measurements
were made using the TE$_{01n}$ resonant modes with a typical
unloaded $Q$ values of $\sim 10^5$ at 10GHz. The temperature of
the dielectric resonator and copper cavity was held constant at
the helium bath temperature. The sample was placed at a magnetic
field antinode    of the dielectric resonator, and was supported
on the end of a long sapphire rod passing centrally through the
resonator. The crystal could be heated from 1.2K to well above
$T_c$  by a heater mounted outside the cavity. Experiments were
performed with the $c$-axis parallel or perpendicular to the rf
magnetic field, to investigate the affect of magnetic anisotropy.
From neutron diffraction experiments,
 the Gd spins are known to align along the $c$-axis \cite{Mook}.
Measurements were made at three of the resonant modes of the
dielectric resonator close to 10, 15 and 21 GHz with suitable
positioning of the sample.

\begin{figure}[tbh]
\begin{center}
\input epsf
\epsfxsize=85mm  \epsfbox{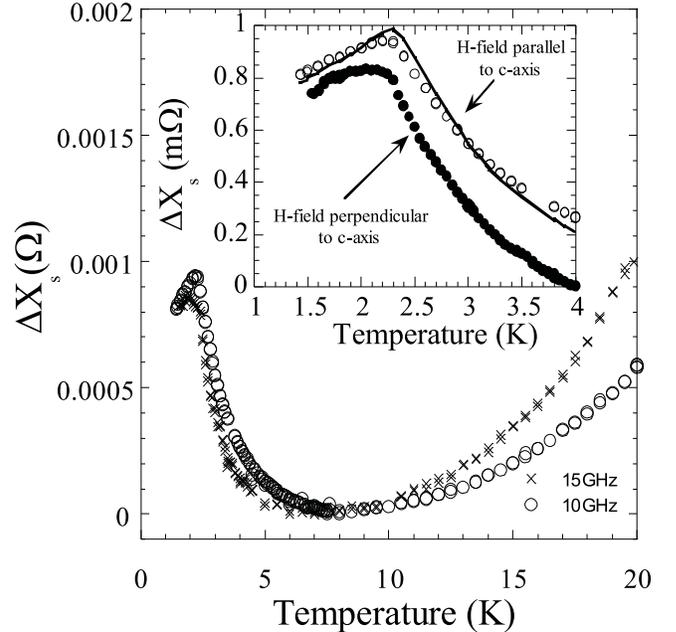} \caption{The surface
reactance at 10 and 15 GHz with the ${\bf H}$-field parallel to
the $c$-axis. The inset shows $\Delta X_s (T)$ for two crystal
orientations at 10GHz, the solid line represents a fit using
equation \ref{fullsigma2}. Open circles are for the ${\bf
H}$-field parallel to the {\it c}-axis and  the closed circles
for the field perpendicular to the $c$-axis.} \label{Xs}
\end{center}
\end{figure}

The microwave properties were determined by a conventional cavity
perturbation method using a HP 8722C Network analyser with
additional data processing to obtain accurate measurements of the
resonant frequency $f_0$ and half power bandwidth $f_B$ of the
dielectric cavity resonances. The changes in these values with
temperature can be related to the surface impedance by the cavity
perturbation formula, \hbox{$\Delta f_B(T) - 2{\rm i}\Delta
f_0(T)= \Gamma (R_s + {\rm i}\Delta X_s)$}. The resonator
constant $\Gamma$ was determined from measurements with the
sample replaced by a chemically polished niobium sample of the
same size and known resistivity. We were able to achieve a
measurement  accuracy and reproducibility for   $R_s$ of $\pm 20
\mu\Omega$ and an error in $\Delta\lambda  = \pm 0.3 {\rm \AA}$,
for a sample of area $0.5 \times 1.1 {\rm mm^2}$ at 10GHz.

\begin{figure}
\begin{center}
\input epsf
\epsfxsize=85mm  \epsfbox{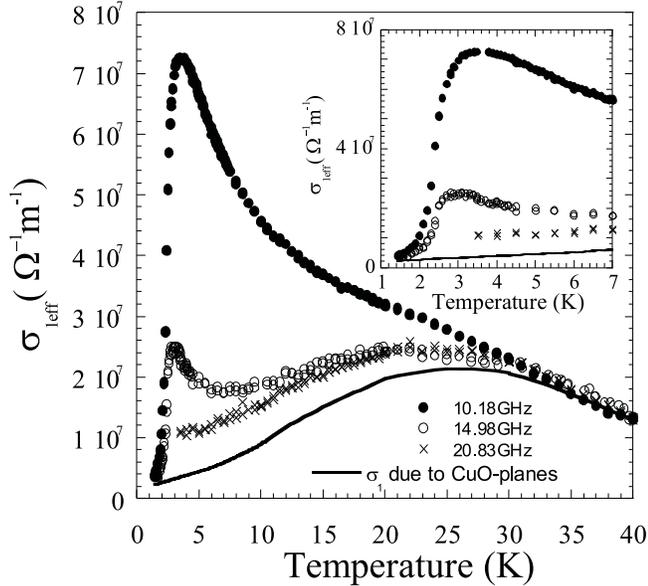} \caption{The
temperature dependence of real part of the microwave conductivity
extracted from the data in figure \ref{Rs}. The solid curve is
$\sigma_1 (T)$ due to the Cu-O planes. The inset is an expanded
view of the low temperature behaviour. } \label{sigma1}
\end{center}
\end{figure}

Figure \ref{Rs} shows measurements of $R_s (T)$ at the three
frequencies. For $T\geq 30$K the losses are quadratic in
$\omega$, consistent with $\omega\tau_{qp}\ll1$. In absolute
terms, the losses are somewhat larger than observed in the best
YBCO crystals, but  are comparable with $R_s$ data for near
optimally doped BSCCO crystals \cite{Wingfield2,Lee}. At lower
temperatures there is a marked and strongly frequency dependent
rise in losses,  which we associate with the paramagnetic
alignment of the Gd spins above the N\'eel temperature. The
losses at 10GHz peak at 3.5K, significantly above the N\'eel
temperature of $2.2 {\rm K}$ (see inset of Figure \ref{Rs}).
These losses then decrease by more than an order of magnitude at
the lowest temperatures.

The spin alignment also leads to a small but pronounced increase
in the reactive part of the surface impedance at low
temperatures, as shown for two frequencies in  figure \ref{Xs}.
This corresponds to an increased penetration depth with a  sharp
change in slope defining the N\'eel temperature. Above $\sim 7$K
, the reactance increases as $\sim T^2$, in contrast to the
$T$-dependence observed for  high quality YBCO crystals. The
magnitude of these losses and the $T^2$ dependence of the
penetration depth imply a larger quasi-particle scattering than
observed in optimally doped YBCO crystals \cite{Bonn}. Above 10K
we have the expected  $\omega$ dependence of $X_s$, but at low
temperatures ($T < 7$K) the reactance has a  $\omega^{-1}$
dependence, as predicted by our model. The inset illustrates
measurements with the crystal aligned with its $c$-axis parallel
and perpendicular to the microwave magnetic field (open and
closed circles respectively). In a parallel microwave field, the
expected direction of spin alignment, the reactance drops
linearly below a well-defined transition temperature $T_N \sim
2.25$K, but in the perpendicular configuration the susceptibility
below $T_N$ is much flatter and may even go through a small
maximum.  Similar results to those illustrated in Figures
\ref{Rs} and \ref{Xs} were observed for all three single crystals
investigated.  The crystals included slightly underdoped,
as-grown and close to optimum-doped oxygen annealed samples, from
different growth batches.
\begin{figure}
\begin{center}
\input epsf
\epsfxsize=85mm  \epsfbox{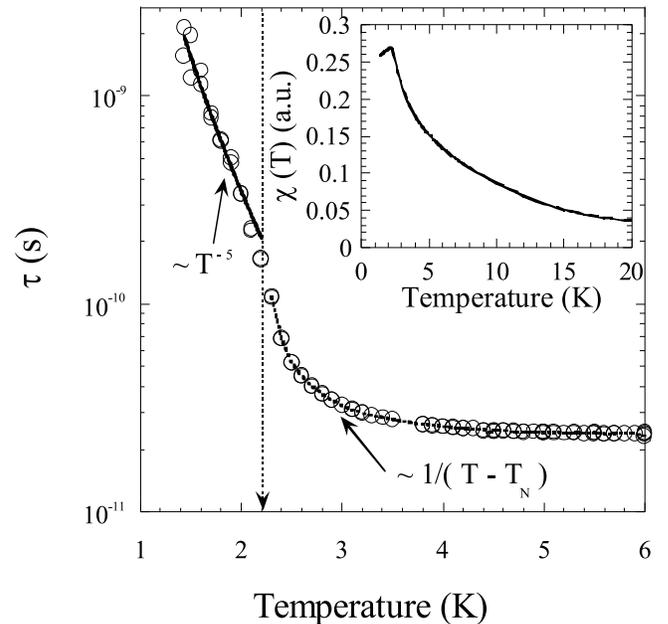} \caption{The
temperature dependence of the spin lattice relaxation time
extracted from the 10 GHz data . At the N\'eel temperature there
is an apparent change in slope in $\tau_s (T)$. Above $T_N$,
$\tau_s (T) \sim [T_N/(T-T_N)]^{\alpha}$ with the best fit
obtained for $\alpha = 1$, below $T_N$ the relaxation time varies
as $\sim T^{-5}$. The inset demonstrates the temperature
dependence of $\chi$ used to extract $\tau_s (T)$. }\label{tau}
\end{center}
\end{figure}

To extract an effective conductivity $\sigma_{1eff}$ from
\hbox{$R_s=\slantfrac{1}{2}
\omega^2\mu_0^2\sigma_{1eff}\lambda^3$}, we have assumed: (i) a
value for $\lambda (0)=140 {\rm nm}$, typical of high quality
YBCO samples (ii) above 10K $\sigma_{2eff}\sim\sigma_2$,  and
(iii) below 10K $\lambda (T)\sim T^2$ consistent with the $T^2$
temperature dependence  of $X_s (T) \sim \omega \mu_0 \lambda
(T)$ in figure \ref{Xs}. We have also assumed that $\sigma_1$ is
not significantly affected by the  Gd spin fluctuations. Figure
\ref{sigma1} shows the temperature dependence of the derived
values of $\sigma_{1eff}$ for all three frequencies measured.
Because the contribution to the effective conductivity from the
Gd spins varies $\sim 1/\omega\tau_s$, using the 10 and 15 GHz
data, we can  extract
 $\sigma_1$ giving the solid line. The derived temperature
dependence  is similar to the  variation of $\sigma_1 ({\rm T})$
observed in YBCO single crystals. It increases to a broad peak at
about 25K, reflecting the increase in the quasi-particle
scattering lifetime at low temperatures.

 In our model, the
additional losses when $\omega\tau_s > 1$ are largely associated
with the paramagnetic relaxation of the Gd spins in the microwave
field. These losses, which we interpret as a decrease in the Gd
spin relaxation rate on approaching and passing through the
antiferromagnetic phase transition, peak at a frequency dependent
temperature significantly above $T_N$. In this respect, we note
that there is no significant change in $\sigma_{1eff}$ at the
N\'eel temperature $T_N$ (see inset of figure \ref{sigma1}). Any
such affect is masked by the much larger changes in $\tau_s$,
figure \ref{tau}.

The region near $T_N$ might be  expected to be dominated by
antiferromagnetic spin fluctuations.  We  assume that close to
$T_N$, $\tau_s$ involves a temperature  dependent term varying as
$[T_N/(T-T_N)]^\alpha$. A fit to this relation for $\alpha = 1$
is  shown in figure \ref{tau}. Below the antiferromagnetic
transition $\tau_s$ increases by an order of magnitude varying
$\sim T^{-5}$ with an apparent  change in slope at $T_N$. The
inset of figure \ref{tau} shows $\chi (T)$  used to derive the
temperature dependence of $\tau_s$. This  assumes  a Curie-Weiss
temperature dependence above $T_N$ and a slight drop in $\chi(T)$
below $T_N$ consistent with the measurement shown in the inset of
figure \ref{Xs}.

The model we have applied assumes that there is no interaction
between the quasi-particles and spins (this is consistent with
specific heat data on GBCO where no change is observed between
semiconducting and superconducting samples \cite{dunlap}).
Susceptibility measurements on non-superconducting GBCO have
shown that $\chi (T)$ fits a 2-d Ising model above $T_N$
\cite{Berg}. Below the transition $\chi (T)$ remains anomalously
high, deviating from the Ising model. This is consistent with our
reactance measurements, where we see only  a small change in
$\Delta X_s (T)$ below $T_N$, figure \ref{Xs}. This is in
contrast to
 what is seen in other RE   substitutions (Sm, Dy and Nd)
where the specific heat data can be fitted to a 2-d Ising model
 both above and below $T_N$ \cite{Maple}.

In  the insets of figures \ref{Rs} and \ref{Xs}, the solid lines
fitted to the data have been evaluated using equations
\ref{fullsigma1} and \ref{fullsigma2}. We have assumed: (i) $\chi
(T)= C/ (T + T_N)$, with a value of $C=n\mu_B^2p^2/12\pi k_B$
corresponding to a derived magnetic moment $p=9.5$, slightly
larger than deduced from magnetic measurements \cite{Thompson}
(ii) the derived Gd spin relaxation time plotted in figure
\ref{tau}, and (iii) the complex electronic conductivity given by
the derived value of $\sigma_1$ in figure \ref{sigma1} and a
value of $\sigma_2$ assuming $\lambda (0) = 140 {\rm nm}$. The
excellent fit to the experimental
   data supports  our  theoretical model. In particular,
 there appears to be no need to invoke any additional effects,
 such as a modification of the electronic mean free path from
 the Gd spin fluctuations.

In summary   we have presented extensive microwave surface
impedance measurements on GBCO single crystals at several
frequencies to investigate the influence of  the
antiferromagnetic alignment  of the Gd spin at low temperatures.
We are able to describe the experimental results by a model
involving  the increase in magnetic  susceptibility associated
with antiferromagnetic alignment and a strongly temperature
dependent relaxation time. The derived  spin lattice relaxation
time increases  below $T_N$ with a temperature dependence $\sim
T^{-5}$. Above $T_N$, $\tau_s \sim 1/(T - T_N)$. Within the
accuracy of our measurements, $\sigma_1 (T)$ is not affected by
the antiferromagnetic alignment of the Gd spins.

We thank G. Walsh  and D. Brewster for valuable technical
support. We also thank   A. Porch and M. Hein for useful
discussions. This research is supported by the EPSRC, UK.


\begin{references}
\bibitem{Hor}P. H. Hor {\it et al.}, Phys. Rev. Lett. {\bf 58}, 1891 (1987).
\bibitem{dunlap}B. D. Dunlap {\it et al.}, Phys. Rev. B {\bf 37}, 592 (1987).
\bibitem{Mook} H. A. Mook  {\it et al.}, Phys. Rev. B {\bf 38}, 12008 (1988).
\bibitem{Waldram} J. R. Waldram, Private communication.
\bibitem{Anlage} S. M. Anlage, Lucia Mercaldo, and  Vladimir Talanov, Private communication.
\bibitem{Wingfield2}  J. J. Wingfield, PhD Thesis, The
University of Birmingham (unpublished) (1999).
\bibitem{Wingfield} J. J. Wingfield, J. R. Powell, C. E. Gough,
and A. Porch, IEEE Trans. Appl. Supercon. {\bf 7}, 2009 (1997).
\bibitem{Thompson} J. R. Thompson {\it et al.}, Phys. Rev. B {\bf 37}, 9395 (1988).
\bibitem{Liang} Ruixing Liang, D. A. Bonn, and W. N. Hardy, Physica C {\bf 304}, 105  (1998).
\bibitem{Erb}A. Erb, E. Walker, and  R. Flukiger, Physica C {\bf 245}, 9  (1996).
\bibitem{twogap} H. Srikanth {\it et al.},
Phys. Rev. B {\bf 55}, R14733 (1997).
\bibitem{Lee}S. F. Lee {\it et al.}, Phys. Rev. Lett. {\bf 77}, 735 (1996).
\bibitem{Bonn} D. A. Bonn {\it et al.}, Phys. Rev. B {\bf 50},
4051 (1994).
\bibitem{Berg} J. van de Berg {\it et al.}, Solid State
Communications {\bf 64}, 699 (1987).
\bibitem{Maple} K. N. Yang {\it et al.}, Phys. Rev. B {\bf 40}, 10963 (1989).
\end{references}
\end{document}